\documentclass[pra,final,tightenlines,showpacs,showkeys,
preprintnumbers,nofootinbib,twocolumn]{revtex4}

\usepackage{graphicx}  
\usepackage{bm}  
\usepackage{amsmath}
\usepackage{epsf}
\newcommand{\beq}{\begin{equation}}
\newcommand{\eeq}{\end{equation}}
\newcommand{\bea}{\begin{eqnarray}}
\newcommand{\eea}{\end{eqnarray}}       

\newcommand{\simlt}{\stackrel{<}{{}_\sim}}

\begin{document}
\preprint{HISKP-TH-06/38}
\preprint{NT@UW-06-29}
\title{Effective Range Corrections to Three-Body Recombination \\ 
for Atoms with Large Scattering Length}
\author{H.-W. Hammer}\email{hammer@itkp.uni-bonn.de}
\affiliation{Helmholtz-Institut f\"ur Strahlen- und Kernphysik (Theorie), \\
Universit\"at Bonn, 53115 Bonn, Germany}
\author{Timo A. L\"ahde}\email{talahde@u.washington.edu}
\affiliation{Department of Physics, University of Washington, \\ 
Seattle, WA 98195-1560, USA}
\author{L. Platter}\email{lplatter@phy.ohiou.edu}
\affiliation{Department of Physics and Astronomy,
Ohio University, Athens, OH 45701, USA\\}
\date{\today}
\begin{abstract}
Few-body systems with large scattering length $a$ have universal 
properties that do not depend on the details of their interactions
at short distances. The rate constant for three-body recombination
of bosonic atoms of mass $m$ into a shallow dimer scales as $\hbar a^4/m$ 
times a log-periodic function of the scattering length. We calculate
the leading and subleading corrections to the rate constant which are
due to the effective range of the atoms and study the correlation
between the rate constant and the atom-dimer scattering length.
Our results are applied to $^4$He atoms as a test case.
\end{abstract}

\maketitle
%
%

\subsection{Introduction}

The properties of low-energy bosons are dominated by S-wave interactions.
If the interaction is short ranged, these properties can
be parametrized by the effective range expansion.
The typical range of the interaction also defines a
natural low-energy length scale $l$ \cite{Braaten:2004rn}.
For a finite range potential $l$ is simply given by the range of the 
potential. If the potential has a van der Waals tail $-C_6/r^6$,
it is determined by the van der Waals length
$l_{vdW}=(mC_6/\hbar^2)^{1/4}$, where $m$ is the mass of the bosons.
For a generic physical system, the effective range parameters, 
such as the scattering length $a$ and the effective
range $r_e$, are of the same order of magnitude as $l$.
In some systems, however, the scattering length $a$ is much
larger than $l$ while the effective range $r_e \sim l$ is still of the 
same order of magnitude. This situation requires a fine tuning of
the parameters in the underlying potential.
Typical examples are $^4$He atoms and nucleons where this
fine tuning is accidental, or alkali atoms near a Feshbach resonance
where it can be controlled experimentally by changing
an external magnetic field. 

Physical systems with large $|a|$ display a number of interesting effects and 
universal properties that are independent of the details of the 
interaction at short distances of order $l$ \cite{Braaten:2004rn}. 
The simplest example for positive $a$ is the existence of a shallow 
two-body bound state (called a dimer) with universal binding energy 
$B_2=\hbar^2/(ma^2)[1+{\cal O}(l/a)]$. Similar relations hold for other 
observables, which can generally be described at low energies by a controlled 
expansion in $l/|a|$. Particularly interesting is the structure of the 
three-boson system.
It has universal properties that include a geometric spectrum of 
three-body bound states (so-called Efimov trimers),
log-periodic dependence of three-body observables on the scattering length,
and a discrete scaling symmetry \cite{Efimov70,Efimov71,Efimov79}. 
These properties can be studied using effective theories which
provide a powerful framework to exploit 
the separation of scales between $a$ and $l$ 
\cite{Braaten:2004rn,Bedaque:1998kg,Bedaque:1998km}.

Since we are mainly interested in applications to atomic systems,
we will refer to the bosons as atoms in the following.
In particular, we focus on three-body recombination,
the process in which three atoms collide to form
a diatomic molecule and an atom. The energy released 
by the binding energy of the molecule goes into the
kinetic energies of the molecule and the recoiling atom.
If the momenta of the incoming atoms are sufficiently small
compared to $1/a$, the momentum dependence of the recombination
rate can be neglected.
The recombination event rate constant $\alpha$ is defined 
such that the number of
recombination events per time and per volume in a gas of cold
atoms with number density $n_A$ is $\alpha n^3_A$. 
If the atom and the dimer produced by the recombination process 
have large enough kinetic energies to escape from the system,
the rate of decrease in the number density of atoms is
\begin{eqnarray}
\frac{d \ }{dt} n_A
& = & - 3 \alpha n^3_A \,.
\label{dnA-gas}
\end{eqnarray}
For large positive $a$, recombination can go into 
the shallow dimer with binding energy $B_2 = \hbar^2/ma^2$ and
another atom. If no deep dimers are present, this is the only
channel for recombination to occur. 
Three-body recombination into deep dimers can also be treated using similar
methods \cite{Braaten:2004rn,BH01,Braaten:2003yc,EGB-99}. 
However, here we assume that no deep dimers
are present. We focus on $a>0$ and consider 3-body
recombination into the shallow dimer. 


\subsection{Theoretical framework}

We first briefly review the universal expressions at leading
order in the expansion in $l/a$.
Dimensional analysis together with the discrete scaling symmetry 
implies that $\alpha$ is proportional to $\hbar a^4/m$ 
with a coefficient that is a log-periodic function of $a$
with period $\pi/s_0$.
To very high accuracy, the three-body recombination constant 
$\alpha$ can be expressed as \cite{NM-99,EGB-99,BBH-00}:
\begin{eqnarray}
\alpha = 67.12\,\sin^2 (s_0 \ln (a/a_{*})+1.67 ) \,
\frac{\hbar a^4}{m} [1+{\cal O}(l/a)]\,,
\label{alpha-eqlo}
\end{eqnarray}
where $s_0=1.00624\ldots$ is a transcendental number and
$a_{*}$ is the scattering length at which there is an Efimov trimer
at the atom-dimer threshold.
Note also that an analytic expression for  
the coefficient 67.12 has recently been derived \cite{Petrov-octs,MOG05}.
The most remarkable feature of the expression 
in Eq.~(\ref{alpha-eqlo}) is that the coefficient of $\hbar a^4/m$ 
oscillates between zero and 67.12 as a function of $\ln(a)$.
In particular, $\alpha$ has zeroes at values of $a$ 
that differ by $e^{\pi/s_0}\approx 22.7$. The locations of these
zeros can be expressed as  $a \approx (e^{\pi/s_0})^n 0.20 \, a_*$
where $n$ is an integer.  The maxima of $\alpha/a^4$  
occur at $a \approx (e^{\pi/s_0})^n 0.94 \, a_*$. These maxima are 
close to the values $a=(e^{\pi/s_0})^n a_*$ for which there is an 
Efimov trimer at the atom-dimer threshold.
In the latter case, atom-dimer scattering becomes resonant and
the atom-dimer scattering length diverges. 

At leading order
in $l/a$, the atom-dimer scattering length can be expressed as
\cite{Efimov79,Braaten:2004rn}:
\beq
a_{ad}=(2.15 \cot (s_0 \ln (a/a_{*}))+1.46)\, a\,[1+{\cal O}(l/a)]\,.
\label{aad-eqlo}
\eeq
Solving Eq.~(\ref{aad-eqlo}) for $a_*$, we can eliminate $a_*$
from Eq.~(\ref{alpha-eqlo}) and obtain an analytical expression
for the universal correlation between $\alpha$ and $a_{ad}$.
In this paper, we study the correlation between
$\alpha$ and $a_{ad}$ up to second order in 
in the expansion in $l/a$. To this order,
these corrections are fully dermined by the S-wave
effective range $r_e$ \cite{Platter:2006ev}. 
Therefore, we refer to them as effective range corrections.

The quantities $\alpha$ and $a_{ad}$ can both be calculated
from the following integral equation:\footnote{For a derivation
of this equation from effective field theory, see e.g.
Ref.~\cite{Braaten:2004rn}.}
\begin{eqnarray}
{\mathcal A}_S(p,k;E) = \frac{16\pi\gamma}{1-\gamma r_e} \: V(p,k;E)
+ \frac{4}{\pi} \int_0^\Lambda \!\!dq \: V(p,q;E) && \nonumber \\
\times \frac{q^2 \,{\mathcal A}_S(q,k;E)}{
-{\gamma} + (\mbox{$\frac{3}{4}$} q^2 -E -i \epsilon )^{1/2} 
- \mbox{$\frac{1}{2}$} r_e (\mbox{$\frac{3}{4}$}q^2-E-\gamma^2)}, \quad &&
\label{BHvK-range}
\end{eqnarray}
where $\Lambda$ is an ultraviolet cutoff, $r_e$ is the effective range, and
$\gamma$ denotes the position of the bound state pole in the atom-atom 
Green's function:
\begin{eqnarray}
\gamma = \left( 1 - \sqrt{1 - 2 r_e/a} \right) \frac{1}{r_e},
\label{gamma-ars}
\end{eqnarray}
where we have set $\hbar=m=1$ for convenience. The potential $V(p,k;E)$ is 
\begin{equation}
V(p,k;E) = \frac{1}{2pk} \ln \left(\frac{p^2 + pk + k^2 -E}
{p^2 - pk + k^2 - E}\right) + G_3(\Lambda),
\end{equation}
where $G_3(\Lambda)$ represents the contribution from the contact three-body 
interaction. For $r_e = 0$, it is given by \cite{Bedaque:1998kg,Bedaque:1998km}
\begin{eqnarray}
G_3(\Lambda)\approx \frac{1}{\Lambda^2} 
\,\cot(s_0 \ln(\Lambda a_*) +2.45)\,,
\end{eqnarray}
but for finite effective range its form is more complicated.
In the following, we will directly calculate the correlation between
$\alpha$ and $a_{ad}$, therefore $a_*$ will not explicitly appear. 
In our case of positive $a$, 
$\gamma^2$ is the binding energy of the shallow dimer.
At leading order (LO) in $l/a$, $\gamma$ reduces to $1/a$ but
the two quantities differ if the effective range is included.
We have chosen  $\gamma$ and $r_e$ as our 2-body inputs
instead of $a$ and $r_e$, because this choice keeps the location 
of the dimer pole fixed which leads to a faster convergence 
of the effective-range expansion. 
The effective theory expansion is then in powers of $\gamma l$.

From the general solution of  Eq.~(\ref{BHvK-range}) one can directly obtain
$\alpha$ and $a_{ad}$.
The three-body recombination constant $\alpha$ is given by 
\beq
\alpha=\frac{8}{\gamma^2\sqrt{3}}\frac{(1-\gamma r_e)}{(1-\gamma r_e/2)^2}
\left| {\mathcal A}_S (0, 2\gamma/\sqrt{3}; 0) \right|^2\,,
\eeq
while the atom-dimer scattering length is given by
\beq
a_{ad}= -\frac{1}{3\pi}\,{\mathcal A}_S (0, 0; -\gamma^2)\,.
\eeq

In principle one can obtain the effective range corrections to all orders 
by solving the integral equation in Eq.~(\ref{BHvK-range}) directly. 
A potential problem comes from the dimer propagator in the second 
line of Eq.~(\ref{BHvK-range}). In addition to the pole from the shallow 
dimer with $\gamma\sim 1/a$, it also has a deep pole
with $\gamma\sim 1/r_e$. This pole is an artefact of the 
form of the dimer propagator and is outside the range of
validity of the effective theory. For negative effective range, 
the pole is on the unphysical sheet of the complex momentum plane
and causes no problems in solving Eq.~(\ref{BHvK-range}).
For positive effective range, the pole is on the physical sheet
and leads to instabilities in the three-body equations
for cutoffs of the order $\sim 1/r_e$ and larger.
We avoid this problem by expanding the kernel of  Eq.~(\ref{BHvK-range})
in the effective range $r_e$ and solving the resulting integral equation.
This allows to calculate the range corrections
perturbatively up to order $(\gamma l)^2$. For higher orders, things become
more complicated and a new three-body parameter enters \cite{Platter:2006ev}.
Moreover, for particles with an interaction displaying
a van der Waals tail $\sim C_6/r^6$, the true interaction can only
be approximated by contact interactions up to order $(\gamma l)^2$.
If a higher accuracy is required, the van der Waals tail has to be taken
into account explicitly \cite{Braaten:2004rn}.
For positive effective range, we perfomed all calculations with a cutoff
well above the breakdown scale of the theory $\Lambda \gg 1/r_e$. For negative
effective range the calculation at next-to-leading order
(NLO) has to be performed with cutoffs
$\Lambda \sim 1/r_e$. This is due to a cancellation between even and uneven
 orders in the expanded two-body propagator at values of the loop
momentum $q \gg 1/r_e$ \cite{Platter:2006ev}. In uneven orders of the 
expansion in $r_e$
this would lead to an incorrect renormalization if the cutoff was
taken much larger than $1/r_e$. In next-to-next-to-leading order
(NNLO), the term proportional
to $(\gamma r_e)^2$ dominates the expansion of the two-body propagator
for large loop momenta and the problem is not present.

We calculate the effective range corrections to the correlation
between $\alpha$ and $a_{ad}$ two different ways: (i) in the 
subtractive renormalization scheme of Ref.~\cite{Platter:2006ev}
where the three-body force term $G_3(\Lambda)$ does not explicitly
appear and (ii) by performing a calculation at fixed cutoff $\Lambda$
and varying $G_3$.\footnote{Since proper renormalization has been
explicitly verified, practical calculations can be carried out at fixed
cutoff $\Lambda$.} 
Both methods agree to very high precision.

\subsection{Correlation between $\alpha$ and $a_{ad}$}

\begin{figure}[t]
\bigskip
\centerline{\includegraphics*[width=8.5cm,angle=0]{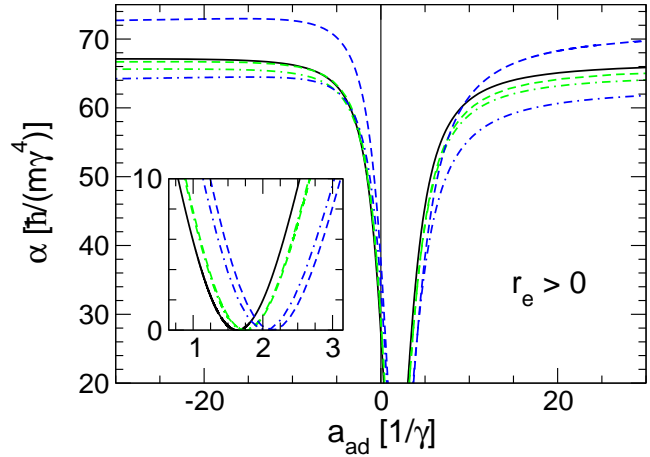}}
\medskip
\caption
{
Correlation between $\alpha$ and $a_{ad}$ for $r_e >0$.
The solid line gives the leading order result, while the dashed (dash-dotted)
lines give NLO (NNLO). The light (dark) curves are for 
$\gamma r_e=0.1$ ($\gamma r_e=0.3$).
}
\label{fig:recp}
\end{figure}

The results for the correlation between $\alpha$ and $a_{ad}$
for positive effective range  $r_e >0$ are shown in Fig.~\ref{fig:recp}
for two different values of the effective range.
The solid line gives the leading order result, while the dashed and dash-dotted
lines give the NLO and NNLO results, respectively. 
The light (dark) curves are for $\gamma r_e=0.1$ ($\gamma r_e=0.3$).
For large absolute values of $a_{ad}$ and $\gamma r_e=0.1$ the shift
from LO to NLO is accidentally very small. However, the shift from 
NLO to NNLO is of the expected size of $(\gamma r_e)^2\sim 0.01$ and
is larger than from LO to NLO. A similar observation holds for the case
$\gamma r_e=0.3$.
We observe that the shift from LO to NLO varies strongly in size and is
sometimes smaller than expected by naive dimensional analysis. However,
the shift from NLO to NNLO is generally of the expected order of magnitude.
The smallness of the NLO corrections can be understood as a cancellation
between two different contributions to this correction.
The two contributions  are proportional to $\gamma r_e$ and $\kappa r_e$,
respectively,  where $\kappa$ is the typical momentum scale
of the process. Furthermore, it is known from experience \cite{Braaten:2004rn}
that at LO observables are often described better than expected from the
power counting once the exact pole position of the two-body propagator is
reproduced. As a consequence, the shifts in observables from LO to NLO
can be very small and of a size comparable to the
corresponding shifts from NLO to NNLO.
The minimum in the correlation curve is moved to larger values of $a_{ad}$.
The size of this shift is of the order $\gamma r_e$ from LO to NLO, and 
of the order $(\gamma r_e)^2$ from NLO to NNLO.

We emphasize that an error estimate determined by the truncation
error of the EFT expansion can be given from dimensional
analysis within the EFT approach.
However, this does not necessarily imply that the shift from one order 
to the next has to be exactly equal to this error estimate.
Firstly, the error estimate is only accurate up to a number of $\mathcal{O}(1)$
which depends on the details of the problem and cannot be estimated
{\it a priori}. Secondly, if there are competing dynamical 
contributions at a given
order, cancellations may lead to significantly smaller corrections as 
can be seen in our results.
  
\begin{figure}[t]
\bigskip
\centerline{\includegraphics*[width=8.5cm,angle=0]{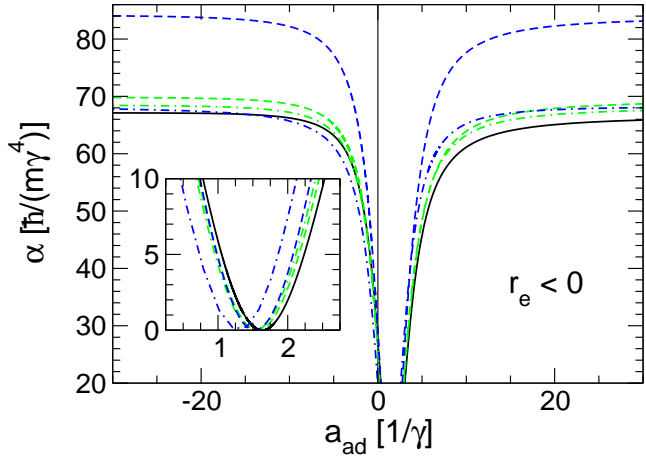}}
\medskip
\caption
{
Correlation between $\alpha$ and $a_{ad}$ for $r_e <0$.
The solid line gives the leading order result, while the dashed (dash-dotted)
lines give NLO (NNLO). The light (dark) curves are for 
$\gamma r_e=-0.1$ ($\gamma r_e=-0.3$).
}
\label{fig:recm}
\end{figure}

The results for the correlation between $\alpha$ and $a_{ad}$
for negative effective range  $r_e <0$ are shown in Fig.~\ref{fig:recm}
for two different values of the effective range. Again,
the solid line gives the leading order result, while the dashed and 
dash-dotted lines give the NLO and NNLO results, respectively. 
The light (dark) curves are for $\gamma r_e=-0.1$ ($\gamma r_e=-0.3$).
For $\gamma r_e=-0.1$, we find that the corrections scale as expected.
In the case $\gamma r_e=-0.3$, the behavior at NNLO resembles the case for 
positive effective range. The minimum is shifted to smaller values of 
$a_{ad}$.  From LO to NLO the size of the corresponding shift
is of the order $\gamma r_e$ and from NLO to NNLO it is
approximately of the order $(\gamma r_e)^2$.
 
For negative effective range, the maximum value of
$\alpha m\gamma^4/\hbar$ in NLO is very sensitive to the value of $r_e$, 
while at NNLO this sensitivity disappears and the maximum value is close 
to the LO result. For $r_e>0$, the sensitivity persists at NNLO.
It is also evident that the convergence of the EFT expansion 
may be slow for large absolute values of $a_{ad}$.
For smaller $|a_{ad}|$, rapid convergence can be achieved for larger 
values of $\gamma r_e$. Indeed it will be shown below that for 
${^4}$He atoms the encountered  values of $\gamma r_e$ are well within 
the range of applicability of the EFT expansion.
The correlation curve has a 
pronounced minimum and is not completely symmetric around this minimum value 
which lies in the interval $1<\gamma a_{ad}^{\rm min}<2$. As a consequence, 
there will always be 2 solutions for $a_{ad}$ if $\alpha$ is used as 
three-body input.

\subsection{Application to $^4$He atoms}

As an illustration, we apply our results to $^4$He atoms
for which the condition $\gamma l\ll 1$ is well satisfied
\cite{Braaten:2004rn,Braaten:2002jv,Braaten:2006vd}. The interatomic potential 
between two $^4$He atoms does not support any deep dimers,
such that our calculations are applicable.
The binding energies of the $^4$He
trimers have been calculated accurately for a number of different model
potentials for the interaction between two $^4$He atoms.
(See, e.g., Refs.~\cite{NFJ98,RY00,MSSK01} and references therein.)
The $^4$He dimer and trimer ground states have been observed in 
experiment \cite{LMKGG,STo96}, 
but the excited state of the trimer has not yet been seen.
Due to the limited experimental information in the three-body sector, 
we apply our effective theory to the TTY~\cite{TTY95}
and HFD-B3-FCI1~\cite{AJM95} potentials.
The input parameters for the effective theory are the dimer binding
energy and the effective range in the two-body sector plus one 
three-body observable. Once these parameters are specified, all
other three-body observables can be predicted to NNLO in the
EFT expansion.
In the following, we use the effective theory to predict $\alpha$
if $a_{ad}$ is known for a given potential and vice versa.

We consider first the TTY potential~\cite{TTY95}.
The scattering length for the TTY potential is $a = 100.01$\ \AA.  
This is much larger than its effective range $r_e = 7.329$\ \AA, 
which is comparable to the van der Waals
length scale $\ell_{\rm vdW} = 5.398$\ \AA~\cite{Braaten:2004rn,Jan00}.  
The binding energy of the $^4$He dimer for the 
TTY potential is $B_2= 1.30962$ mK
which is small compared to the natural low-energy 
scale for $^4$He atoms: $E_{\rm vdW} \approx 420$~mK. 
This energy $B_2$ corresponds to a pole position 
$\gamma=0.01040$\ \AA$^{-1}$. 
The atom-dimer scattering length for the TTY potential
was calculated in Ref.~\cite{Ro03}
as $a_{ad}= (1.205 \pm 0.001)\gamma^{-1}$. Using this information
as the three-body input and the effective range for the TTY potential,
$r_e \gamma =0.076$,
we can obtain the recombination constant $\alpha$ which has not been
calculated for the TTY potential. Our results are given in Table \ref{tab:He1}.

\begin{table}[h]
\caption{
Recombination constant $\alpha$ in units of $\hbar/(m\gamma^4)$
for the TTY potential \cite{TTY95}. The errors given correspond
to the uncertainty in the input quantity $a_{ad}$.
(Note that $\hbar^2/m=12.1194$ K\AA$^2$
for $^4$He atoms.)}
\vspace*{0.3cm}
\begin{tabular}{l||c|c|c}
              & LO & NLO & NNLO\\ 
\hline 
$\alpha\ [\hbar/(m\gamma^4)]$  & $\:2.791 \pm 0.013\:$ 
                               & $\:3.899 \pm 0.015\:$ 
                               & $\:3.778 \pm 0.014\:$ \\
\end{tabular}
\label{tab:He1}
\end{table}

The values are numerically accurate to the number of digits given.
Even at NNLO the numerical errors in the input parameters 
are typically negligible compared to the truncation error of the effective 
theory. As an illustration, we have propagated the numerical error in the
input parameter $a_{ad}$ to the results in Table~\ref{tab:He1}.
It should be emphasized that the physical accuracy is determined by the 
truncation error of the 
effective theory.  The effective theory error estimates
are 10\%, 1\%, and 0.1\% times a number
of $\mathcal{O}(1)$ at LO, NLO and NNLO, respectively.
The actual difference between LO and NLO and NLO and NNLO in our
results is 40\% and 4\%, respectively,
but still in agreement with the above
expectation. From this pattern we expect our NNLO result to be 
accurate to 0.4\%. This slightly larger error than expected
is due to the value of $\alpha m\gamma^4/\hbar$
close to the minimum in the correlation curve, where small
corrections are relatively more important.
In Ref.~\cite{Braaten:2002jv}, $\alpha$ was calculated at LO
using the trimer excited state energy instead of $a_{ad}$
as the three-body input and the value $\alpha=2.9\,\hbar /(m\gamma^4)$ 
was found. This result is in good agreement with our value at LO.

For the HFD-B3-FCI1 potential~\cite{AJM95}, the situation is the
opposite and $a_{ad}$ is unknown. However, the recombination constant 
$\alpha=12\cdot 10^{-29}$ cm$^6$/s has been calculated~\cite{SEGB02}.
The scattering length for this potential is $a = 91.0$\ \AA, 
which is again much larger than the effective range $r_e = 7.291$\ \AA\,
and the van der Waals
length scale $\ell_{\rm vdW} = 5.398$\ \AA~\cite{Braaten:2004rn,Jan00}. 
The position of the dimer pole is $\gamma=0.01149$\ \AA$^{-1}$.
From this information, we can extract $r_e \gamma = 0.084$
and $\alpha m\gamma^4/\hbar=1.32$. Because of the shape of the 
curve in Fig.~\ref{fig:recp}, we have now two solutions for
$a_{ad}$ as shown in Table~\ref{tab:He2}.

\begin{table}[h]
\caption{
The two solutions for the atom-dimer scattering 
length $a_{ad}$ in units of $\gamma^{-1}$
for the HFD-B3-FCI1  potential \cite{AJM95}.
(Note that $\hbar^2/m=12.1194$ K\AA$^2$
for $^4$He atoms.)}
\vspace*{0.3cm}
\begin{tabular}{l||c|c|c}
              & LO & NLO & NNLO \\ 
\hline 
$a_{ad} [\gamma^{-1} ]$  (solution 1) & \:1.341\: & \:1.434\: & \:1.424\: \\
$a_{ad} [\gamma^{-1} ]$  (solution 2) & \:1.931\: & \:2.034\: & \:2.024\: \\
\end{tabular}
\label{tab:He2}
\end{table}

Using the information on $a_{ad}$ from the TTY potential above,
we can identify solution 1 as the appropriate one for $^4$He atoms.
The effective theory error estimates
are 10\%, 1\%, and 0.1\% times a number of $\mathcal{O}(1)$
at LO, NLO, and NNLO, respectively.
In this case, the difference between LO and NLO as compared to
the difference between NLO and NNLO is
a factor of 0.7 smaller than the above estimate and thus
in good agreement with the expectation. From this
pattern we expect the error of our NNLO result to be
0.1 \%.
In Ref.~\cite{Braaten:2004rn}, $a_{ad}$ was calculated at LO
using the trimer excited state energy instead of $\alpha$
as the three-body input and the value 
$\gamma a_{ad}=1.4$ was found.
Again, this result is in good agreement with our value at LO.
The deviation between results using different three-body inputs
gives a simple estimate of higher order corrections. For $a_{ad}$,
this estimate works well. In the calculation of $\alpha$
discussed above, however, 
this estimate is misleading because the value of $\alpha$
is close to the minimum of the correlation curve.

\subsection{Conclusions}

In this work we have computed the effective range correction to the 
recombination coefficient $\alpha$ to second order in the 
expansion in $\gamma l$. Up to this order, only the two-body
scattering length $a$ and effective range $r_e$ plus one 
three-body parameter are required as input for the effective
theory \cite{Platter:2006ev}. We have calculated the
correlation between the atom-dimer scattering length $a_{ad}$ and the 
recombination coefficient $\alpha$ at LO, NLO and NNLO, and studied the 
convergence of the effective theory expansion in $\gamma l$. 
We found that the convergence of the EFT expansion 
may be slow for large absolute values of $a_{ad}$.
For natural values $|\gamma a_{ad}|\simlt 3$, 
rapid convergence could be achieved for 
$\gamma r_e \simlt 0.3$.  In the same 
manner, the correlation between $\alpha$ and other low-energy three-body
observables, such as Efimov trimer energies, could be calculated.
As a test, we have applied our results to $^4$He atoms.  We have used the
TTY \cite{TTY95} and HFD-B3-FCI1 potentials \cite{AJM95} as input and 
predicted, using the effective theory, three-body observables that have 
not yet been calculated for these potentials. The convergence of the expansion 
in $\gamma l$ for $^4$He atoms was found to be in agreement with
the {\it a priori} expectation.

Further efforts should be devoted to the inclusion of effective range 
corrections for recombination
processes in ultracold gases of alkali atoms.
Such atoms have many deep two-body bound states which modify 
the recombination into the shallow dimer and provide additional 
channels for recombination into the deep states.
At LO in the expansion in $\gamma l$ these effects have already been
calculated in this effective theory 
\cite{Braaten:2004rn,BH01,Braaten:2003yc,Braaten:2006vd}.
An extension to NNLO and the inclusion of
temperature dependence (see Refs.~\cite{Jonsell06,YFT06,Braaten:2006nn})
should allow for a precise description of
experimental loss rates such as obtained by the Innsbruck 
group for $^{133}$Cs atoms \cite{Grimm06}.
The recombination of cold helium atoms at nonzero collision
energies has already been considered by Suno and 
coworkers \cite{Suno02}.
A further step to increase the precision with which these experiments
can be described theoretically is to include the recombination into
Efimov trimers and dimer-dimer scattering resonances \cite{Chin05,FB18Santos}.
Recent results in the four-body sector give hope that this goal can be
achieved within the present effective theory 
framework~\cite{Platter:2004qn,Yama06,Hammer:2006ct}.
\begin{acknowledgments}
This research was supported by the U.S. Department of Energy under 
grant DE-FG02-93ER40756 (LP) and DE-FG02-97ER41014 (TL), and by an Ohio 
University postdoctoral fellowship. We thank D.R.~Phillips for discussions. 
LP thanks the theory group of the HISKP at Bonn University for its hospitality 
during the completion of this work.
\end{acknowledgments}
%
%
%

\end{document}